# BENNETT CLOCKING OF NANOMAGNETIC LOGIC USING ELECTRICALLY INDUCED ROTATION OF MAGNETIZATION IN MULTIFERROIC SINGLE-DOMAIN NANOMAGNETS

J. Atulasimha[1,(a)] and S. Bandyopadhyay[(b)]

(a)Department of Mechanical Engineering

(b)Department of Electrical and Computer Engineering

Virginia Commonwealth University

Richmond, VA 23284, USA

## Abstract

The authors show that it is possible to rotate the magnetization of a multiferroic (strain-coupled two-layer magnetostrictive-piezoelectric) nanomagnet by a large angle with a small electrostatic potential. This can implement Bennett clocking in nanomagnetic logic arrays resulting in unidirectional propagation of logic bits from one stage to another. This method of Bennett clocking is superior to using spin-transfer torque or local magnetic fields for magnetization rotation. For realistic parameters, it is shown that a potential of ~ 0.2 V applied to a multiferroic nanomagnet can rotate its magnetization by nearly $90^0$ to implement Bennett clocking.

---

[1] Corresponding author. E-mail: jatulasimha@vcu.edu

Nanomagnetic logic – also known as magnetic quantum cellular automata [1] – is an energy-efficient computing paradigm that can work at room temperature [1-6]. In this architecture, classical binary bits 0 and 1 are encoded in two stable magnetization orientations parallel and anti-parallel to the easy axis of magnetization of a nanomagnet with large shape anisotropy. Logic gates are configured by exploiting the dipole-dipole interaction between nearest neighbor nanomagnets [1, 2].

In order to propagate logic bits unidirectionally from an input to an output stage down a chain of nanomagnets, one normally requires a clock signal to periodically reset the magnetization direction of each nanomagnet. This can be done either with a global magnetic field which reorients the magnetization of *all* the nanomagnets along the hard axis at the same time [2], or with a local agent [3] that reorients the magnetization of every nanomagnet independently. The former scheme does not allow pipelining of data and the magnetization of a magnet along the hard axis must be maintained until a bit has propagated through it. This requires generating a local energy minimum around the hard axis, using, for example, materials with biaxial anisotropy [7], but this energy minimum is typically so shallow that thermal noise can quickly relax the magnetization to the easy axis resulting in large bit error probability [8]. Therefore, we will address only the local clocking scheme which is free of these shortcomings.

The most effective local clocking scheme is due to Bennett [9]. To understand how it works, consider a nanomagnet array in the ground state shown in the first row of Fig. 1. In this state, the magnetizations of nearest neighbors are anti-parallel owing to dipole-dipole interaction (Fig. 1: first row). As a result, the logic bit encoded in the magnetization orientation of the first nanomagnet is replicated in every odd-numbered nanomagnet. Therefore, this array acts like a wire (or a series of delay-gates) to transmit the first bit down the chain.

If we flip the magnetization of the first nanomagnet to switch its bit state (Fig. 1: second row), we expect all succeeding nanomagnets to flip in a domino effect so that the new bit state is propagated down the chain. However, this may not happen; the second magnet's magnetization does not necessarily flip since the second magnet finds its left neighbor telling it to flip, while its right neighbor (still in its original state) forbids flipping. Since both influences are equally strong, the array is stuck in a metastable state and the logic wire fails. To break this logjam, one forcibly turns the magnetizations of the second and third nanomagnet by a large angle with a local agent acting as a "clock" (Fig. 1: third row). When this agent is finally removed from the second nanomagnet, the latter finds itself in an asymmetric environment (left neighbor magnetized along easy axis and right neighbor close to the hard axis) which allows it to flip its magnetization (Fig. 1: fourth row) and reach the lowest energy state which will be the desired logic state. Thus, by sequentially rotating the magnetizations of magnet-pairs through a large angle (with a multi-phase clock), one can propagate the new state of the first nanomagnet (input logic bit) unidirectionally down the chain. This is the essence of Bennett clocking.

In this letter, we show that if a nanomagnet is composed of a multiferroic material [10-11], then its magnetization *can be rotated through large angles [12] by the stress generated in the magnetostrictive layer when the piezoelectric layer is subjected to an electrostatic potential*. We also show that this can propagate logic bits unidirectionally and implement Bennet clocking.

Consider a chain of single-domain multiferroic nanomagnets as shown in Fig. 1. The total energy of any nanomagnet in this chain, subjected to a stress generated by an electrostatic potential, is $E_{total} = E_{dipole} + E_{shape-anisotropy} + E_{stress-anisotropy}$, where $E_{dipole}$ is the dipole-dipole interaction energy due to interaction with nearest neighbors, $E_{shape-anisotropy}$ is the shape anisotropy energy due to the magnets anisotropic shape, and $E_{stress-anisotropy}$ is the stress anisotropy energy caused by the electrostatic potential.

We now focus on the second and third nanomagnets which are clocked (subjected to electrostatic potentials) and assume that their magnetizations subtend angles $\theta_2$ and $\theta_3$ with the positive x-axis. The dipole interaction energy of the second nanomagnet can be expressed as [13]:

$$E_{dipole} = -\mu_0 \vec{M} \bullet \vec{H}_{dipole} V = \left(\mu_0/4\pi R^3\right)\left[M_s^2 V^2\right]\left[-2\cos\theta_2 \cos\theta_3 + \left(\sin\theta_3 - 1\right)\sin\theta_2\right] \quad (1)$$

where $\mu_0$ is the permeability of free space, $\vec{H}_{dipole}$ is the in-plane magnetic field due to a dipole, and $V$ is the volume of the nanomagnet.

The shape anisotropy energy (energy difference between magnetization along the hard and the easy axis) is given by [13]:

$$E_{shape-anisotropy} = \left(\mu_0/2\right)\left[M_s^2 V\right] N_d \quad (2)$$

where $N_d$ is the demagnetization factor. To obtain analytical expressions for the demagnetization factors, we treat the elliptical nanomagnet as a flat ellipsoid in the manner discussed by Chikazumi [13]. Its major and minor axis diameters are $a$ and $b$, while the thickness is $t$. For the co-ordinate system consistent with Fig 1, the expressions for $N_d$ along the $y$ (major axis) and $x$ (minor axis) directions are respectively [13]:

$$N_{d\_YY} = \frac{\pi}{4}\left(\frac{t}{a}\right)\left[1 + \frac{1}{4}\left(\frac{a-b}{a}\right) - \frac{3}{16}\left(\frac{a-b}{a}\right)^2\right]; \; N_{d\_XX} = \frac{\pi}{4}\left(\frac{t}{a}\right)\left[1 + \frac{5}{4}\left(\frac{a-b}{a}\right) + \frac{21}{16}\left(\frac{a-b}{a}\right)^2\right], (3)$$

provided $a>b$, $a/b\sim1$ and $a, b >t$ [13].

Since the magnetization of the second nanomagnet subtends an angle $\theta_2$ with the positive x-axis, the shape anisotropy energy can be written as

$$E_{shape-anisotropy} = \frac{\mu_0}{2}\left[M_s^2 V\right]\left(N_{d\_XX}\cos^2\theta_2 + N_{d\_YY}\sin^2\theta_2\right) = \frac{\mu_0}{2}\left[M_s^2 V\right]\left(N_{d\_YY} + \{N_{d\_XX} - N_{d\_YY}\}\cos^2\theta_2\right). \quad (4)$$

Finally, the stress anisotropy energy is given by [13]:

$$E_{stress-anisotropy} = -\frac{3}{2}\left[\lambda_s \sigma V\right]\sin^2\theta_2, \quad (5)$$

where $(3/2)\lambda_s$ is the saturation magnetostriction. Compressive stress will make $\sigma$ negative and tensile positive.

Using equations (1) - (5), we can write the total energy of the second nanomagnet in Fig 1 as

$$\begin{aligned}E_{total-2} &= \left(\mu_0/4\pi R^3\right)\left[M_s^2 V^2\right]\left[-2\cos\theta_3\cos\theta_2 + (\sin\theta_3 - 1)\sin\theta_2\right] \\ &+ \left(\mu_0/2\right)\left[M_s^2 V\right]\left(N_{d\_XX} - N_{d\_YY}\right)\cos^2\theta_2 - (3/2)\lambda_s \sigma V \sin^2\theta_2,\end{aligned} \quad (6)$$

and similarly the total energy of the third nanomagnet is

$$\begin{aligned}E_{total-3} &= \left(\mu_0/4\pi R^3\right)\left[M_s^2 V^2\right]\left[-2\cos\theta_2\cos\theta_3 + (\sin\theta_2 - 1)\sin\theta_3\right] \\ &+ \left(\mu_0/2\right)\left[M_s^2 V\right]\left(N_{d\_XX} - N_{d\_YY}\right)\cos^2\theta_3 - (3/2)\lambda_s \sigma V \sin^2\theta_3,\end{aligned} \quad (7)$$

where we have dropped constant terms which do not depend on $\theta_2$ and/or $\theta_3$.

In order to demonstrate that the magnetizations of the second and third nanomagnets indeed rotate upon application of stress and that the former subsequently settles down in the correct logic state when it is unstressed [indicating that signal has propagated unidirectionally], we have to solve equations (6) and (7) simultaneously to find the energy minima of both nanomagnets as a function of stress applied, with the appropriate initial conditions. An accurate transient solution will require solving the Landau-Lifshitz-Gilbert (LLG) equations for the coupled system, but the final steady-state solution does not require it. All we have to do is minimize both $E_{total-2}$ and $E_{total-3}$ as we gradually increase and decrease stress on the nanomagnets and find the values of $\theta_2$ and $\theta_3$ corresponding to the energy minima at each value of the stress. This calculation is carried out numerically.

For the numerical simulation, the multiferroic nanomagnets were assumed to be made of nickel/lead-zirconate-titanate (PZT) with the following properties: for nickel layer, thickness = 10 nm, $(3/2)\lambda_s = -3 \times 10^{-5}$, $M_s = 4.84 \times 10^5$ A/m [14], and Young's modulus $Y = 2 \times 10^{11}$ Pa. The PZT layer layer [15, 16] can transfer up to $500 \times 10^{-6}$ strain to the Ni. The major and minor axes of the elliptical nanomagnets are assumed to be $a = 105$ nm and $b = 95$ nm respectively and the center-to-center separation (or pitch) is 160 nm. The above parameters were chosen to ensure that: (i) The shape anisotropy energy of the nanomagnets is sufficiently high (~0.8 eV or ~32$kT$ at room temperature) so that the bit error probability due to spontaneous magnetization flipping is very low ($\sim e^{-32} \approx 10^{-14}$); (ii) The stress anisotropy energy (~1.5 eV) generated in the magnetostrictive Ni layer due to a strain of $500 \times 10^{-6}$ transferred from the PZT layer can rotate the second magnet out of the initial state so that upon removal of the stress, it flips to the correct state as shown in Fig 1; (iii) The dipole interaction energy is limited to 0.2 eV which is significantly lower than the shape anisotropy energy. This prevents the magnetization from switching spontaneously without the application of the electric-field induced stress.

Equations (6) and (7) are evaluated as functions of orientations $\theta_2$ and $\theta_3$ to find the energy minima. Initial conditions are σ=0, $\theta_2$=-90º and $\theta_3$=+90º corresponding to the top row of Fig 1. For each increment in stress, the new $\theta_2$ and $\theta_3$ are simultaneously evaluated. Furthermore, it is assumed that both magnetization orientations rotate to the right to simplify the numerical analysis. The analysis would be identical if both magnetizations rotated to the left, because of the symmetry. We first rotate the second nanomagnet by applying stress to it and then rotate the third nanomagnet. This ensures that the probability of one magnet rotating to the left and the other rotating to the right is very small because the x-component of $\vec{H}_{dipole}$ on the third nanomagnet exceeds the y-component when the second nanomagnet has already completed its rotation through a large angle. The x-component favors lining up the rotated magnetizations in the same direction (parallel) while the y-component favors lining them up in opposite directions (anti-parallel).

In Fig. 2(a), we plot $\theta_2$ and $\theta_3$ as a function of tensile stress applied on the second nanomagnet while the third nanomagnet is left unstressed. Surprisingly, there is no rotation in both $\theta_2$ and $\theta_3$ for small values of stress. But at a certain threshold stress [~58 MPa], the second magnet switches abruptly from $\theta_2 = -90^0$ to $\theta_2 = +6^0$ while the third nanomagnet switches abruptly from $\theta_3 = +90^0$ to $\theta_3 = +75^0$. Even though the third magnet is unstressed, it rotates because the second magnet's rotation away from $-90^0$ immediately causes an x-component of $\vec{H}_{dipole}$ to appear on the third magnet which then makes it rotate. Further increase in stress on the second magnet has little effect on $\theta_3$ but $\theta_2$ decreases asymptotically to $0^0$ when the stress is increased to 100 MPa.

In Fig. 2(b), we plot $\theta_2$ and $\theta_3$ when stress is held constant at 100 MPa on the second nanomagnet and gradually increased on the third nanomagnet. There is no abrupt switching of magnetization. Gradually $\theta_3$ decreases from $+75^0$ to $+6^0$ while $\theta_2$ increases from close to $0^0$ to $+6^0$. At this point, the initial anti-parallel configuration of the second and third nanomagnet has become parallel as one magnetization has rotated clockwise and the other anti-clockwise (both have rotated to the right) to reach a final state of $\theta_2 = \theta_3 = 6^0$.

Finally, in Fig. 2(c) we plot these angles as the stress is gradually reduced to zero on the second nanomagnet while being held constant at 100 MPa on the third nanomagnet. The second nanomagnet rotates into the "nearly-up" state $\left(\theta_2 \approx 80^0\right)$ while the third nanomagnet aligns approximately along the hard axis $\left(\theta_3 \approx 0.2^0\right)$. Thus, at the end of the stress cycle on the second nanomagnet, the latter has flipped from its initial "down" state to the "up" state, i.e. it has successfully switched. *Repeating this sequence on the next pair of nanomagnets (third and fourth) propagates the input logic bit (magnetization orientation of the first nanomagnet) down the chain.* Thus, Bennett clocking is successfully implemented.

We also investigated the cases when stress was applied simultaneously on both the second and the third nanomagnets, as well as sequentially on the third followed by the second [see supplementary material]. The dynamics of magnetization rotations are different in these cases, but in the end, when both magnets are stressed at 100 MPa, their magnetizations always assume the same orientations $\theta_2 = \theta_3 = 6^0$. Therefore, the final results are identical in all cases.

A deeper physical understanding of the magnetization rotation can be gained by looking at the energy profiles of the second and third nanomagnets as a function of $\theta_2$ and $\theta_3$ (Figs. S3, S4 and S5 in the supplementary material) at zero and other intermediate stresses. These profiles can be found in the supplementary material.

In conclusion, we have shown that nanomagnets can be electrically switched for Bennett clocking. Using the piezoelectric coefficient of PZT ($d_{31}$ ~ -100×10$^{-12}$ m/V), a voltage of 0.2 V will be required to induce a stress of 100 MPa in a 40 nm PZT layer, assuming linear behavior. This method of switching is superior to using a local magnetic field to rotate the magnetic moment since a magnetic field cannot be confined easily to dimensions of ~ 100 nm (cell size) and cannot be generated easily. It is also superior to the use of spin transfer torque [5] since that typically requires a spin-polarized current which is not easily produced.

**Figure Captions**

**Figure 1:** Logic propagation with Bennett clocking using multiferroic nanomagnets. (First row) a chain of elliptical nanomagnets in the ground state with magnetization orientation indicated by arrows. (Second row) Magnetization of the first magnet is flipped down with an external agent and the second magnet finds itself in a tied state where its left neighbor's dipole interaction favors flipping but its right neighbor's dipole interaction does not. (Third row) The second and the third magnet are subjected to electrically induced stress that rotates their magnetizations. (Fourth row) The second magnet is freed from stress and it switches to the desired "up" state since the dipole interaction from the left neighbor is now stronger than that from the right neighbor so that the tie is resolved. The right panel shows the energy landscape of the second magnet corresponding to the rows.

**Figure 2:** Rotation angle $\theta_2$ of the second nanomagnet and $\theta_3$ of the third nanomagnet as a function of the stress. (a) For increasing stress applied to the second nanomagnet while third nanomagnet is not stressed (b) for increasing stress applied to the third nanomagnet while a constant stress of 100 MPa is maintained on the second nanomagnet (c) for decreasing stress on the second nanomagnet to zero while a constant stress of 100 MPa is maintained on the third nanomagnet.

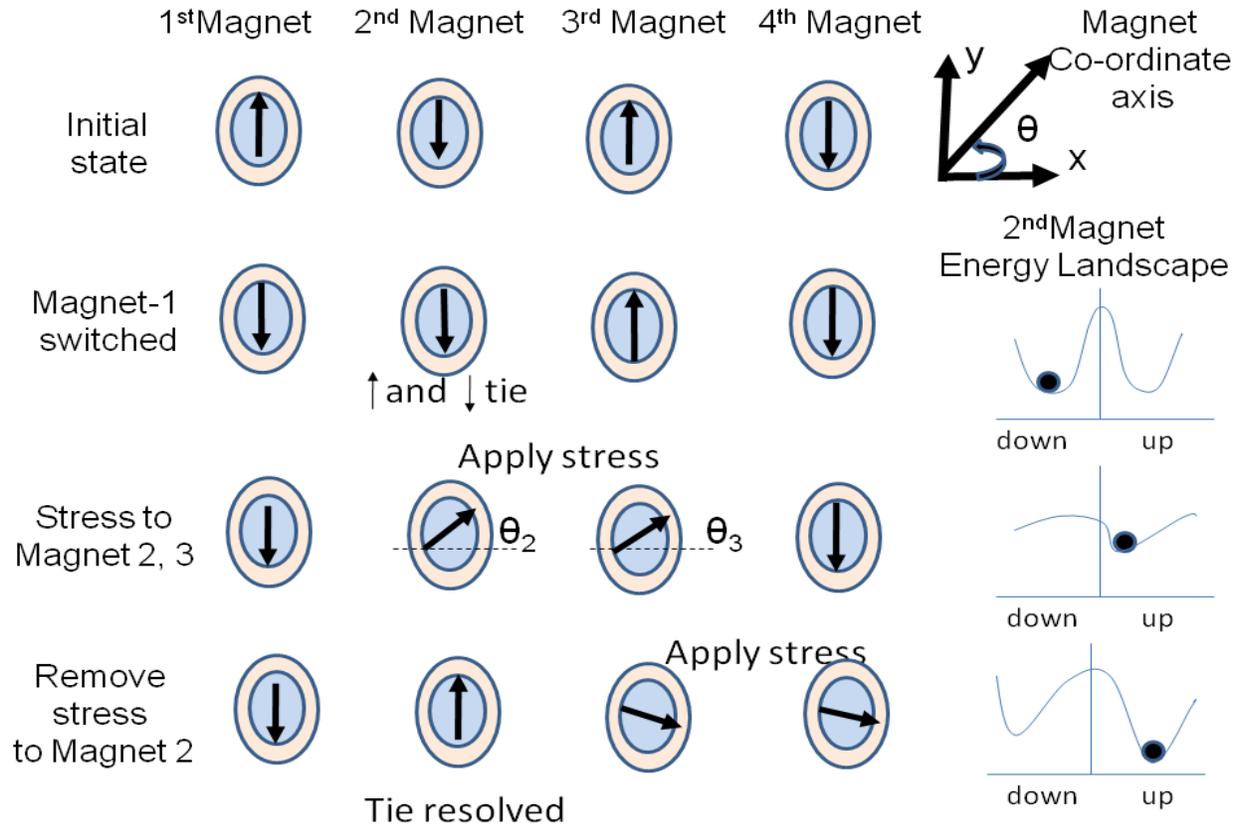

Fig. 1

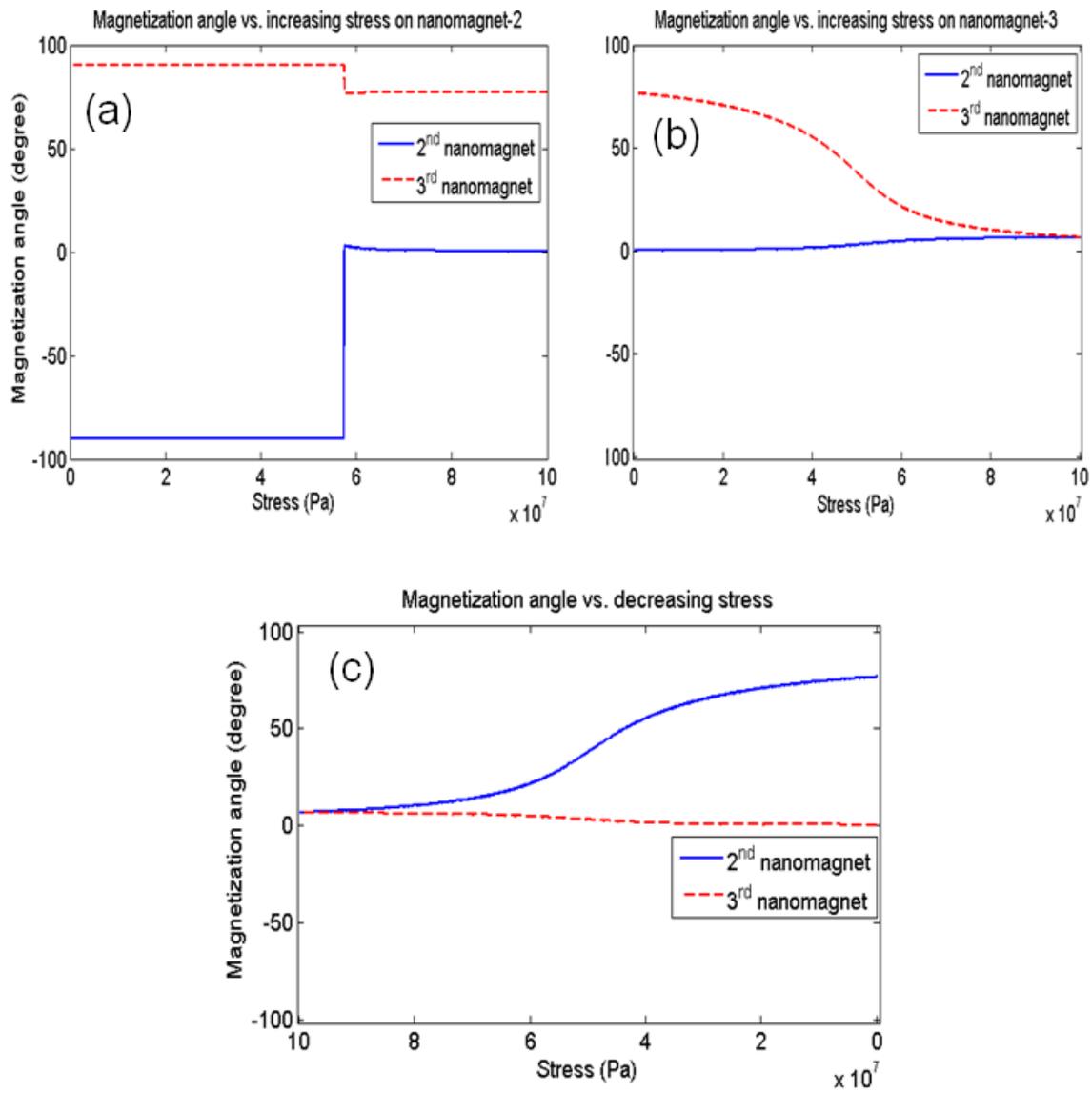

Fig. 2

**SUPPLEMENTARY MATERIAL TO ACCOMPANY "BENNETT CLOCKING OF NANOMAGNETIC LOGIC USING ELECTRICALLY INDUCED ROTATION OF MAGNETIZATION IN MULTIFERROIC SINGLE-DOMAIN NANOMAGNETS"**

**Magnetization rotation under different sequences of stressing**

We discussed the magnetization rotations that take place when stress is applied first on the second nanomagnet and then on the third. Here, we discuss the remaining two situations, namely (i) when stress is applied simultaneously on both the second and the third nanomagnets, and (ii) when stress is applied first on the third and then on the second nanomagnet.

I. Simultaneous stressing of both nanomagnets

We start with the same initial conditions $\sigma=0$, $\theta_2=-90°$ and $\theta_3=90°$ corresponding to the top row of Fig 1. Both the second and the third nanomagnet are then subjected to tensile stress *at the same time*. We assume that both magnetization orientations will rotate to the right. As before, the analysis would be identical if both magnetizations rotated to the left, because of the symmetry. However, in this case, there is actually a much larger probability that one magnet will rotate to the right and the other to the left because there is initially no x-component of $\vec{H}_{dipole}$ acting on either magnet. It is this x-component that tends to keep the two magnetizations parallel as opposed to anti-parallel. Therefore, in order to make both magnets rotate to the right or left, they should be stressed sequentially so that as soon as one rotates, an x-component of $\vec{H}_{dipole}$ appears on the other, which will attempt to align it parallel with its partner. Consequently, simultaneous stressing is not desirable.

In Fig S1, we plot $\theta_2$ and $\theta_3$ as a function of tensile stress applied simultaneously on both the second and the third nanomagnet. Once again, there is no change in either $\theta_2$ or $\theta_3$ (no rotation) for small values of stress, but at a certain threshold stress (~58 MPa) an *abrupt* switching occurs from $\theta_2 = -90^0$ to $\theta_2 \approx +17^0$ in the second nanomagnet while the third nanomagnet simultaneously switches from $\theta_3 = +90^0$ to $\theta_3 \approx +17^0$. Thus, once again, the anti-parallel configuration suddenly becomes parallel as one magnetization rotates clockwise and the other anti-clockwise (both rotating to the right). Further increase in stress rotates both magnetizations asymptotically toward $\theta_2 = \theta_3 = 0^0$, saturating at about $\theta_2 = \theta_3 = 6^0$ at 100 MPa. This is the same orientation as that achieved when stress is applied first on the second nanomagnet and then on the third. Therefore, the final orientations are the same in both cases.

Next, as we gradually decrease the stress to zero on the second nanomagnet while holding it constant at 100 MPa on the third nanomagnet, both $\theta_2$ and $\theta_3$ will obviously replicate the behavior shown in Fig. 2 (c). Therefore, after the end of the stress cycle on the second nanomagnet, the latter's logic state has flipped from the initial "down' state to the desired "up" state. At this point, the input logic bit has propagated through the first two magnets. Repeating this sequence will make it propagate through the succeeding magnets.

II. <u>Stress is applied first on the third magnet and then on the second while holding the third at a constant stress</u>

In Fig. S2 (a), we show the magnetization orientations of both the third and the second magnets when only the third magnet is stressed. No rotation occurs in either magnet until the stress on the third exceeds 72 MPa. After that, the third magnet rotates abruptly from $\theta_3 = +90^0$ to $\theta_3 = 50^0$ while the second rotates from $\theta_2 = -90^0$ to $\theta_2 = -74^0$. As soon as the magnetization of the third nanomagnet rotates away from +90° to the right, an x-component of $\vec{H}_{dipole}$ appears on the second magnet which makes it rotate to the right even though it is unstressed. Then, as the stress is increased further to 100 MPa on the third magnet, it rotates gradually to $\theta_3 = 19^0$ while the second rotates slightly to $\theta_2 = -70^0$.

Next, when stress is applied on the second nanomagnet, while the third is held at the maximum stress of 100 MPa (Fig S2 (b)), the magnetization of the second nanomagnet gradually rotates away from $\theta_2 = -74^0$ to about $\theta_2 = -55^0$. At a stress of ~ 26 MPa on the second nanomagnet, its orientation suddenly switches from $\theta_2 = -55^0$ to $\theta_2 = +67^0$. At the same time, the magnetization of the third nanomagnet rotates abruptly from $\theta_3 = 19^0$ to $\theta_3 \approx 0^0$. Thereafter, further stressing the second magnet up to 100 MPa rotates the magnetizations of both magnets to finally reach the state $\theta_2 = \theta_3 = 6^0$. Note that at 100 MPa stress, the final magnetization orientations of both magnets are the same as before, i.e. they are always the same regardless of whether stress is applied simultaneously on them or sequentially.

Finally, as stress on the second nanomagnet is gradually relaxed while holding the stress on the third magnet constant at 100 MPa, the angles $\theta_2$ and $\theta_3$ will replicate the behaviors shown in Fig. 2 (c). Therefore, once again, the logic state of the second nanomagnet would flip from the initial "down" state to the desired "up" state at the end of its stress cycle.

A few interesting observations can be made at this point. First, rendering the magnetizations of the second and third magnets parallel starting from their initial anti-parallel configuration requires less stress (58 MPa) when both magnets are stressed simultaneously than when they are stressed sequentially (100 MPa). Second, rotating the magnetization of the third nanomagnet away from its initial $+90^0$ angle with stress is much harder than rotating the magnetization of the second magnet away from its initial $-90^0$ orientation. This is because in the former case, both the shape anisotropy energy and the dipole energies due to the second and fourth nanomagnet pointing down (Fig 1) have to be overcome, while in the latter case, only the shape anisotropy energy has to be overcome. The dipole energy is absent in the latter case since the first nanomagnet points down while the third points up.

**ENERGY PLOTS**

The magnetization rotations in Figs. 2, S1 and S2 can be understood better by examining the energy profiles of the second and third nanomagnets as a function of $\theta_2$ and $\theta_3$ at zero and other intermediate stresses. We will plot the energy profiles for three cases:

Case I: Stress is applied first on the second nanomagnet, gradually increased to 100 MPa, and then stress is applied on the third nanomagnet up to 100 MPa while holding the second nanomagnet's stress constant at 100 MPa. Subsequently, the stress on second nanomagnet is reduced to zero while the third nanomagnet is held at a stress of 100 MPa

In Figures S3 (a) and S3 (b), we plot the energy profiles of the second and third nanomagnets as functions of $\theta_2$ and $\theta_3$ respectively, as the second magnet is increasingly stressed and the third magnet is left unstressed. The energy profile of the second magnet is always symmetric about $\theta_2 = 0$ since this magnet experiences no net dipole interaction (one of its neighbors is pointing up and the other pointing down). However, that is not the case for the third nanomagnet since it experiences strong dipole interaction owing to the fact that both its neighbors are initially pointing in the same direction. Therefore, this magnet's energy profile is not symmetric about $\theta_3 = 0$ but is tilted towards positive values of $\theta_3$.

With increasing stress on the second nanomagnet (going from curve "1" towards curve "2" in Fig. S3 (a)), the barrier separating the minima in magnet 2 is gradually eroded, but the locations of the minima do not change resulting in no rotation of magnetization. At a critical stress, the stress anisotropy energy overcomes the shape anisotropy energy (past the curve marked "2") and the barrier vanishes completely. At this point, the second nanomagnet's energy minima move abruptly from $\theta_2 = \pm 90^0$ to $\theta_2 = +6^0$ as shown in the curve marked "3" (the two global minima collapse into a single global minimum at a different location). As soon as the second magnet has thus switched, the third magnet's global energy minimum moves abruptly from $\theta_3 = +90^0$ to $\theta_3 = +75^0$ because it suddenly experiences the x-component of $\vec{H}_{dipole}$ which appears when the second nanomagnet switches. This is shown in Fig. S3 (b). Further increase in stress on the second magnet moves its own minimum's location closer to $\theta_2 = 0^0$, but does not affect the third nanomagnet's energy minimum which stays at $\theta_3 = +75^0$. Note that the third magnet's energy profile is practically independent of the stress applied to the second magnet until it switches. After the switching event, the energy profile of the third magnet changes, but the new profile is again relatively unresponsive to stress on the second magnet. Therefore, the third magnet has two stable energy profiles.

Figures S3 (c) and S3 (d) show the energy profiles of the second and third magnet as functions of $\theta_2$ and $\theta_3$ respectively, when increasing stress is applied on the third nanomagnet, while holding the second nanomagnet's stress constant at 100 MPa. The second magnet's minimum moves back from $\theta_2 = 0°$ to $\theta_2 = +6°$ as it experiences changing dipole interaction due to the rotation of the third magnet, while the energy minimum of the third nanomagnet gradually moves from $\theta_3 = +75°$ to $\theta_3 = +6°$ with increasing stress on it. Note that at some intermediate stress (curve marked "3" in Fig. S3 (d)), the double minima structure of the third magnet's energy profile transforms into a single minimum structure. Since the dipole fields generated by the first and fourth nanomagnets point upwards, the magnetizations of the second and third nanomagnets can never reach the orientations $\theta_2 = \theta_3 = 0°$ (exact alignment along the hard axis) at any finite stress, but can only approach these orientations asymptotically.

Figures S3 (e) and S3 (f) show the energy profiles of the second and third nanomagnets as functions of $\theta_2$ and $\theta_3$ respectively when the stress on the second nanomagnet is gradually decreased to zero, while holding the third nanomagnet at 100 MPa constant stress. Profiles marked "1", "2" and "3" show decreasing stress from "1" through "3". Initially, the single minimum of the second nanomagnet shifts to larger $\theta_2$ when stress is lowered (curve marked "2"). Then, with further decrease in stress, two minima states appear, but the one at positive $\theta_2$ is ~ 0.25 eV (10 kT at room temperature) lower than the one at negative $\theta_2$ ensuring that the former is the global minimum. Therefore, the second magnet's moment settles into the "up-state" $(\theta_2 \to 90°)$ even for a non-quasi-static (fast) switching.

One issue that needs some discussion is that whenever there are two minima at some value of stress in either nanomagnet and the barrier separating them is of the order of ~ kT, it is possible to thermally transition between these minima by absorbing phonons. Such transitions require large rotations or magnetic reversals which typically require ~ 1 nanosecond to complete. In order to suppress this process, stress must be applied rapidly enough to reach from 0 to 100 MPa in less than 1 nanosecond. With the assumed Young's modulus of nickel, the strain generated in the nanomagnet is ~$10^{-3}$, so that a 100 nm long magnet has to expand by ~ 1 Å in ~ 1 nanosecond. We see no fundamental barriers to this taking place.

Case II: Stress is applied simultaneously on both the second and the third nanomagnet

As in Case I, the energy minima are at $\theta = \pm 90^0$ in both nanomagnets owing to shape anisotropy. With increasing stress (going from curve marked "1" towards curve marked "2" in Figs. S4 (a) and S4 (b)) the barrier separating the minima is reduced, but as long as the locations of the minima do not change, there is no rotation of magnetization. At a critical stress, when the stress anisotropy energy overcomes the shape anisotropy energy (past the curve marked "2"), the barrier vanishes and the second nanomagnet's energy minimum moves abruptly from $\theta_2 = -90^0$ close to $\theta_2 = 17^0$ as shown in the curve marked "3". Once the second magnet has thus switched, the third magnet's energy minimum also moves from $\theta_3 = +90^0$ close to $\theta_3 = 17^0$ (curve marked "3"), so that this magnet also switches abruptly. As in case I, the magnetizations of the second and third nanomagnets can never reach the orientations $\theta_2 = \theta_3 = 0^0$ (exact alignment along the hard axis) at any finite stress, but can only approach these orientations asymptotically.

Case III: Stress is applied first on the third nanomagnet, gradually increased to 100 MPa, and then stress is applied on the second nanomagnet while holding the third nanomagnet's stress constant at 100 MPa

Once again, the initial minima in both nanomagnets are at $\theta = \pm 90^0$ because of shape anisotropy. When stress is applied to the third nanomagnet while leaving the second magnet unstressed, the asymmetric barrier separating the minima in the third nanomagnet [caused by both shape anisotropy and dipole interaction energy from second and fourth nanomagnet] is gradually eroded (Fig. S 5 (b)). Upon reaching a stress of 72 MPa, the barrier vanishes and the third magnet's energy minimum moves abruptly from $\theta_3 = +90^0$ to $\theta_3 = 50^0$ (curves "2" to "3" in Fig S5 (b)). This causes sudden rotation of the third magnet's magnetization from $+90^0$ to $+50^0$.

At this point, the second magnet's energy minimum moves from $\theta_2 = -90^0$ to $\theta_2 = -74^0$ (curves "2" to "3" in Fig S5 (a)) as the symmetry of this magnet's energy profile about $\theta_2 = 0^0$ is destroyed by the x-component of $\vec{H}_{dipole}$ that suddenly appears as the third magnet rotates abruptly from $\theta_3 = +90^0$ to $\theta_3 = 50^0$. A local minimum now develops at $-74^0$ and the second magnet rotates to this angle from $-90^0$. Interestingly, the global minimum of the second magnet occurs at $+74^0$, but there is a large energy barrier of ~ 0.3 eV that prevents the magnet from transitioning to that state.

Then, as the stress on the third magnet is increased to 100 MPa, its energy minimum gradually moves to $\theta_3 = 19^0$ [curve marked "4" in Fig. S5 (b)] as increasing stress anisotropy favors orientations away from 90º and closer to $0^0$. The second nanomagnet magnetization rotates only slightly from $\theta_2 = -74^0$ to $\theta_2 = -70^0$ as no stress is applied on it and the only distortion to its energy profile is caused by the dipole interaction energy from the magnetization rotation in the third nanomagnet.

When stress is applied on the second nanomagnet, with the third nanomagnet held at the maximum stress of 100 MPa, the key switching events are: (i) at a stress of ~ 26 MPa, the barrier due to shape anisotropy is completely overcome by the stress anisotropy and the second magnet's energy minimum switches suddenly from $\theta_2 = -55^0$ to $\theta_2 = +67^0$ (curves "2" and "3" in Fig S5 (c)). This event reduces the net y-component of $\vec{H}_{dipole}$ appearing on the third nanomagnet due to dipole interaction with nanomagnets 2 and 4. As a result, the energy minimum of the third nanomagnet moves abruptly from $\theta_3 = 19^0$ to $\theta_3 \approx 0^0$ (curves "2" and "3" in Fig S5 (d)). Further increase in stress asymptotically moves the energy minima in both magnets to the orientations $\theta_2 = \theta_3 = 6^0$.

SUPPLEMENT FIGURE CAPTIONS

**Figure S1:** Rotation angle $\theta_2$ of the second nanomagnet and $\theta_3$ of the third nanomagnet as a function of increasing stress applied simultaneously to both nanomagnets.

**Figure S2:** Rotation angle $\theta_2$ of the second nanomagnet and $\theta_3$ of the third nanomagnet (a) for increasing stress on the third nanomagnet while the second nanomagnet is unstressed, (b) for increasing stress on the second nanomagnet while the third is maintained at a constant stress of 100 MPa.

**Figure S3:** Energy landscape of the nanomagnets with stress sequentially on the second and then the third nanomagnet. (a) Energy $E_{total-2}$ of the second nanomagnet as a function of $\theta_2$ and (b) energy $E_{total-3}$ of the third nanomagnet as a function of $\theta_3$, for increasing stress on second nanomagnet while the third nanomagnet is unstressed. (c) Energy $E_{total-2}$ of the second nanomagnet as a function of $\theta_2$ and (d) energy $E_{total-3}$ of the third nanomagnet as a function of $\theta_3$, for increasing stress on the third nanomagnet while stress on second nanomagnet is maintained at 100 MPa. (e) Energy $E_{total-2}$ of the second nanomagnet as a function of $\theta_2$ and (f) energy $E_{total-3}$ of the third nanomagnet as a function of $\theta_3$ with decreasing stress on the second nanomagnet, while the stress on the third nanomagnet is held constant at 100 MPa. The numbers "1", "2", "3" in each figure indicate sequence of increasing stress in (a) to (d), and decreasing stress in (e) and (f)).

**Figure S4:** Energy landscape of the nanomagnets when stress is applied on both at the same time. (a) Energy $E_{total-2}$ of the second nanomagnet as a function of $\theta_2$ and (b) energy $E_{total-3}$ of the third nanomagnet as a function of $\theta_3$ for increasing stress applied simultaneously on both nanomagnets. The numbers "1", "2", "3" in each figure indicate sequence of increasing stress.

**Figure S5:** Energy landscape of the nanomagnets with stress applied sequentially on the third and then the second nanomagnet. (a) Energy $E_{total-2}$ of the second nanomagnet as a function of $\theta_2$ and (b) energy $E_{total-3}$ of the third nanomagnet as a function of $\theta_3$ for increasing stress applied on the third nanomagnet, while the second is unstressed. (c) Energy $E_{total-2}$ of the second nanomagnet as a function of $\theta_2$ and (d) energy $E_{total-3}$ of the third nanomagnet as a function of $\theta_3$ for increasing stress applied on the second nanomagnet, while the third is at 100 MPa stress. The numbers "1", "2", "3" in each figure indicate sequence of increasing stress.

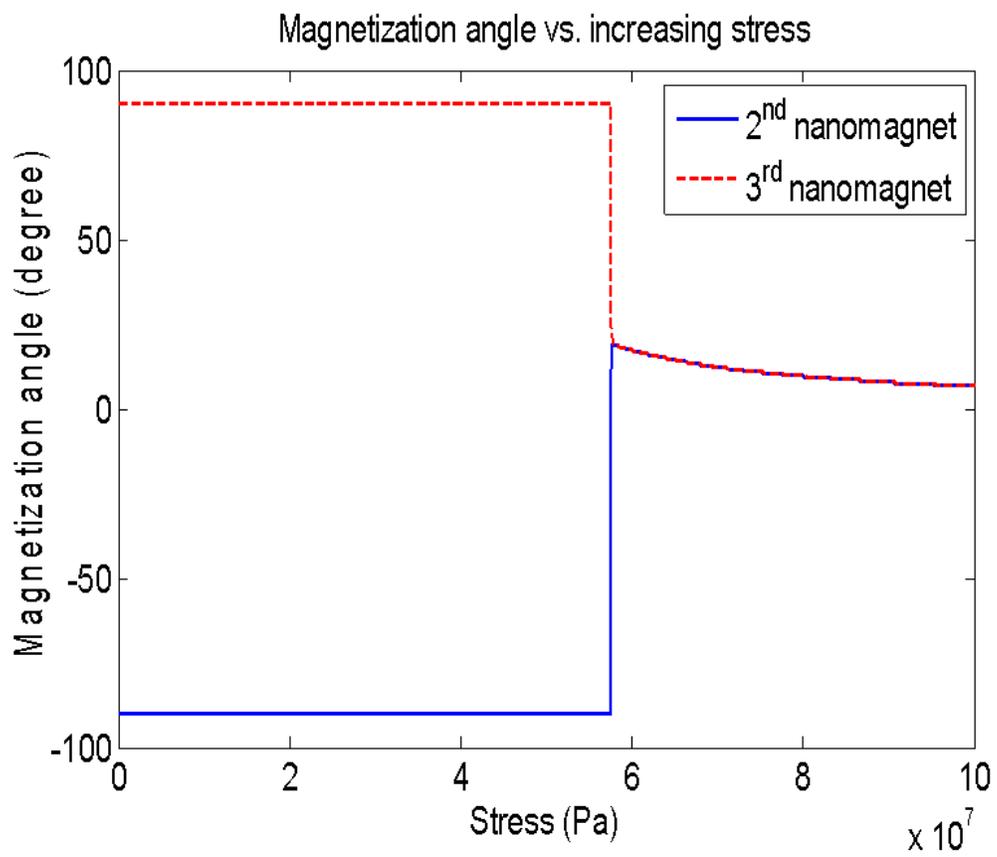

Fig. S1

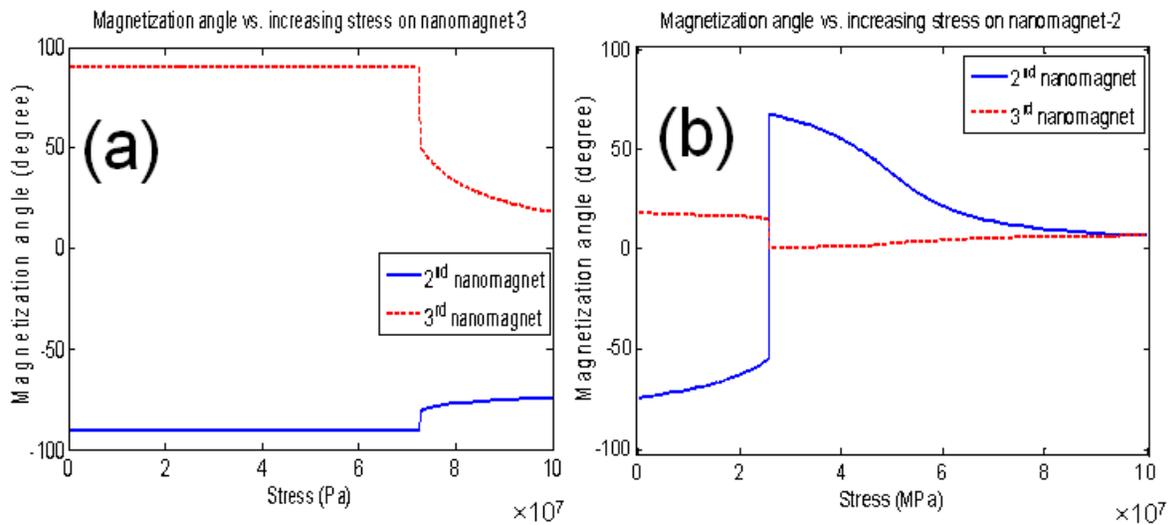

Fig. S2

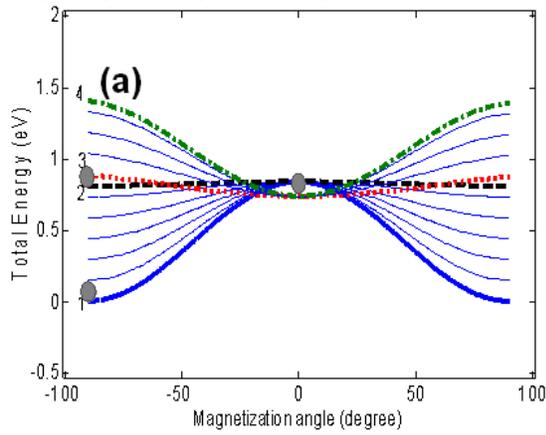
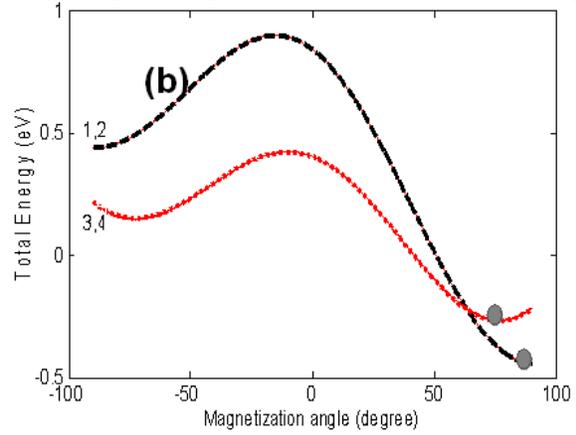
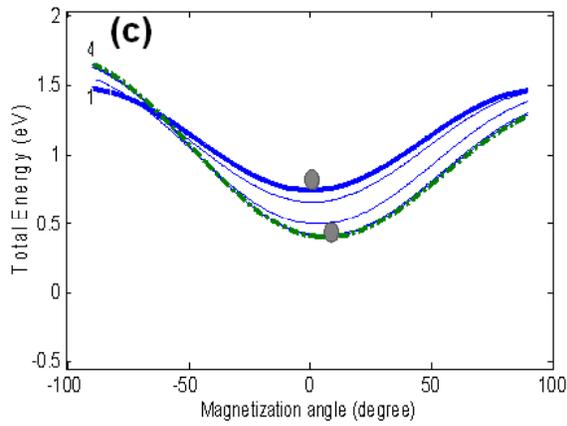
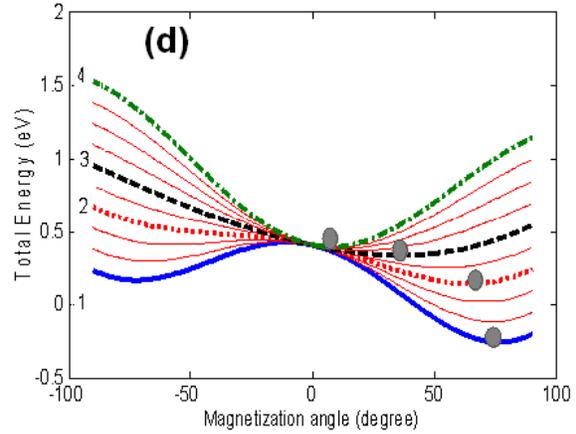
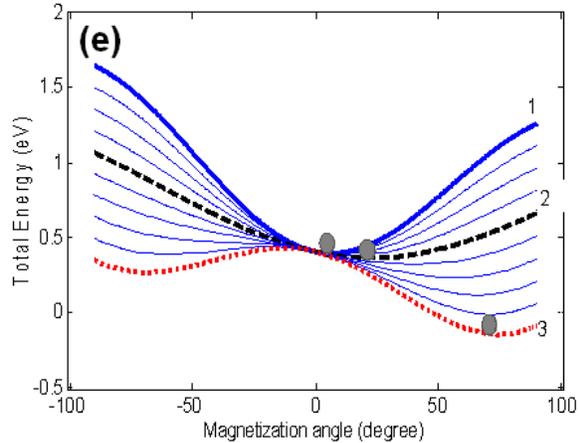
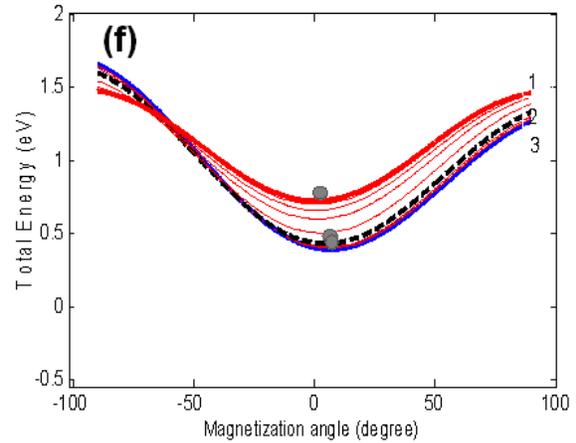

Fig. S3

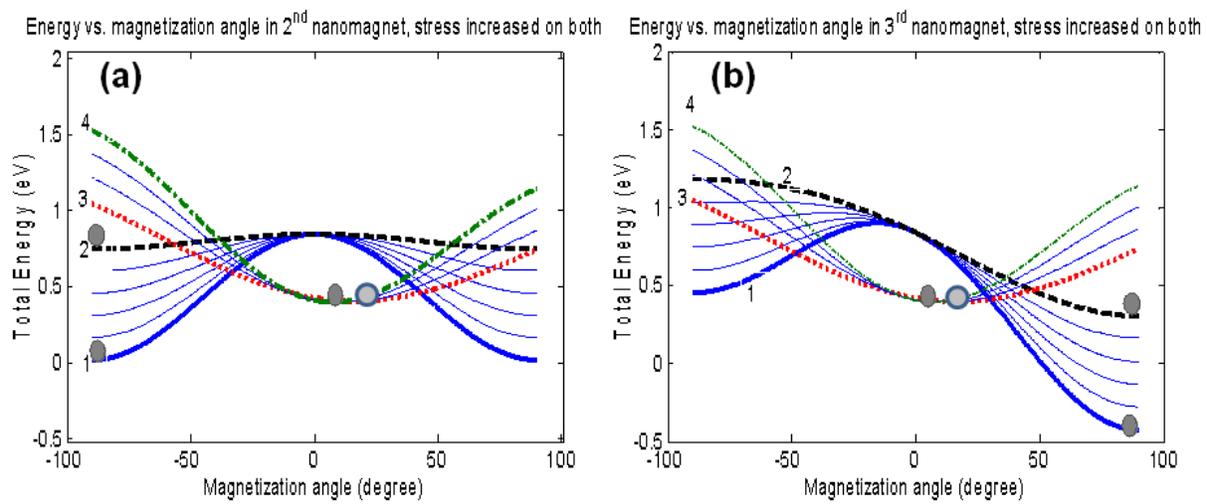

Fig. S4

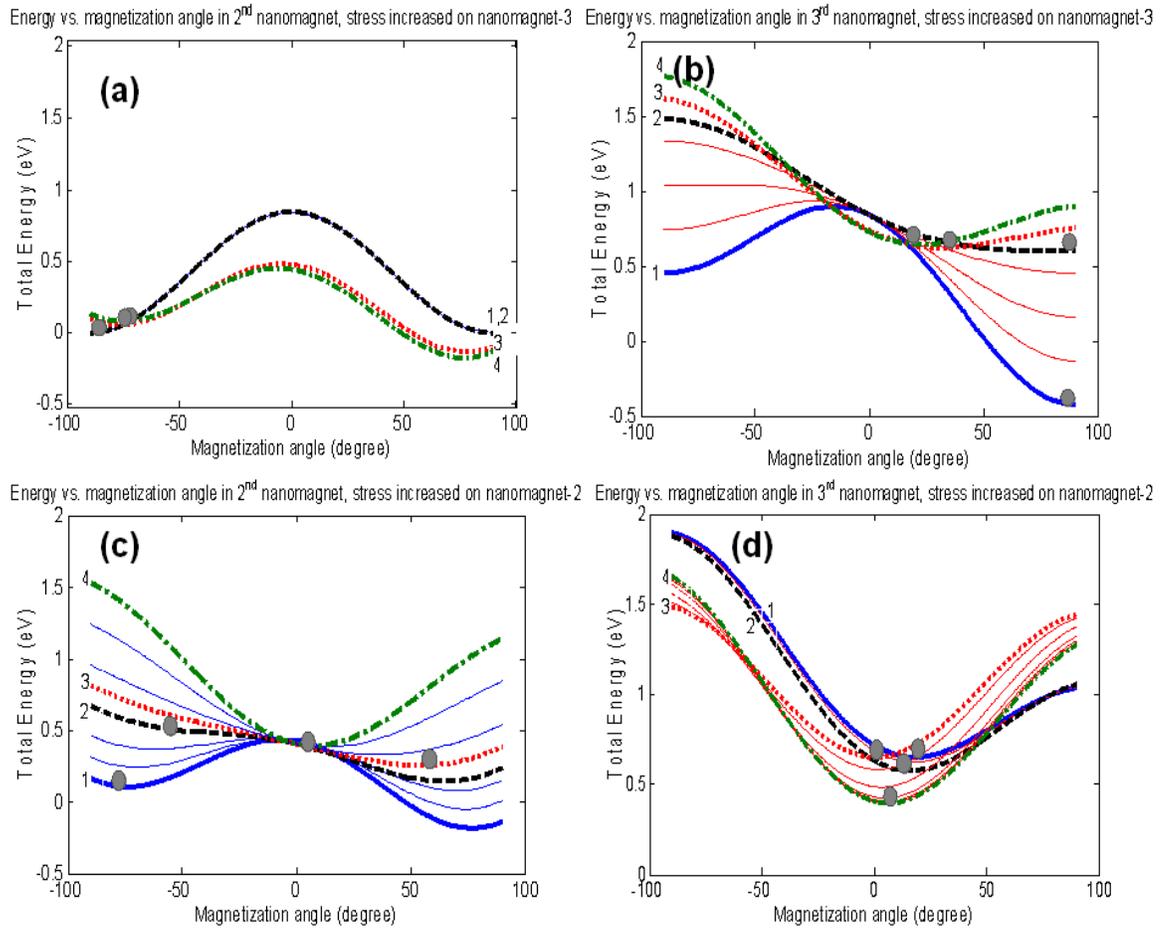

Fig. S5